\documentclass[twocolumn,showpacs,preprintnumbers,amsmath,amssymb]{revtex4}
\usepackage{graphicx}
\usepackage{dcolumn}
\usepackage{bm}
\usepackage{natbib}
\begin{document}

\title{Topological Catastrophe and Isostructural Phase Transition in Calcium}
\author{Travis E. Jones$^\dagger$}
 \email{trjones@mines.edu}
\author{Mark E. Eberhart$^\dagger$}
\author{Dennis P. Clougherty$^{\dagger\ddagger}$}

\affiliation{$^\dagger$ Molecular Theory Group,
                         Colorado School of Mines,
                         Golden, Colorado 80401 }

\affiliation{$^\ddagger$ Department of Physics,
                   University of Vermont,
                   Burlington, Vermont 05405-0125 }

\date{\today}

\begin{abstract}
We predict a quantum phase transition in fcc Ca under hydrostatic pressure.   Using density functional theory, we find at pressures below 80 kbar, the topology of the electron charge density is characterized by nearest neighbor atoms connected through bifurcated bond paths and deep minima in the octahedral holes. At pressures above 80 kbar, the atoms bond through non-nuclear maxima that form in the octahedral holes. This topological change in the charge density softens the C$'$ elastic modulus of fcc Ca, while C$_{44}$ remains unchanged. We propose an order parameter based on applying Morse theory to the charge density, and we show that near the critical point it follows the expected mean-field scaling law with reduced pressure.
\end{abstract}

\maketitle

The theory and characterization of solid-solid isostructural phase transitions has been an area of intense experimental and theoretical research since they were described in 1949 \cite{Lawson1949}. What sets these transitions apart from others is that while crystalline properties change--sometimes discontinuously--crystallographic structure does not, making these transitions purely electronic in nature. Both first order and continuous transitions are known to exist. A well known example of the former is the $\gamma \rightarrow \alpha$ transition in fcc Ce \cite{Lawson1949}, while a magneto-volume instability in fcc Fe is an example of the later \cite{Kong:2004}. 

In the case of Fe, the traditional local order parameter, $\mu$, gave conflicting results computationally, with the transition appearing as first \cite{Moruzzi:1989} or second \cite{Kong:2004} order. It was not until the transition was interpreted as a  topological change of the elctronic charge density, $\rho(\vec{r})$, that the underlying nature of the phase change became apparent \cite{Jones}. While this approach is promising, no general correlations have been drawn between isostructural phase transitions and $\rho(\vec{r})$.  We use this approach to predict an isostructural phase transition in calcium under hydrostatic pressure.  We then apply Morse theory to the charge density and propose an order parameter for the transition.  We show that this order parameter scales in the appropriate way with reduced pressure near the quantum critical point.  

One of the advantages of casting theories of isostructural phase transition in terms of $\rho(\vec{r})$ is that it is a quantum observable. Though $\rho(\vec{r})$ is often calculated by way of first principles methods (e.g.~density functional theory (DFT)), the total charge density can also be measured via X-ray diffraction techniques, and the spin-polarized charge density can be determined using spin-polarized neutron diffraction. As it is also  known  that all ground state properties depend on the charge density \cite{HK}, it seems appropriate to seek relationships between the structure of  $\rho(\vec{r})$, changes to that structure, and corresponding changes to properties.  We note that previous work has proposed that DFT could be used as a general framework for analyzing quantum phase transitions in electronic systems \cite{qc-dft}.  

The electron charge density is a 3 dimensional scalar field. Morse theory tells us the charge density can be characterized by its rank 3 critical points (CPs), the points at which the gradient of the field vanishes. There are four kinds of rank 3 CP in a three-dimensional space: a local minimum, a local maximum and two kinds of saddle point. These are conventionally denoted by an index, which is the number of positive curvatures minus the number of negative curvatures. For example, a minimum CP has positive curvature in three orthogonal directions; therefore it is called a (3, +3) CP. The first number is the dimensionality of the space, and the second number is the net number of positive curvatures. A maximum is denoted by (3, -3), since all three curvatures are negative. A saddle point with two of the three curvatures negative is denoted (3, -1), while the other saddle point is a (3, +1) CP.

Through extensive studies of molecules and crystals, Bader and Zou showed that it was possible to correlate topological properties of $\rho(\vec{r})$ with elements of molecular structure and bonding \cite{AIM,Zou:1994}. In particular, a bond path was shown to correlate with the ridge of maximum charge density connecting two nuclei. The existence of such a ridge is guaranteed by the presence of a (3, -1) CP. As such, this CP is often referred to as a bond CP.  Other types of CPs have been correlated with other features of molecular structure. A (3, +1) CP is topologically required at the center of ring structures. Accordingly, it is designated a ring CP. Cage structures must enclose a single (3, +3) CP and are given the name cage CPs.  The locations of the atomic nuclei always coincide with a (3, -3) CP. Hence, it is often called an atom or nuclear CP.  
 
In this study, the charge density of fcc Ca was calculated at various lattice constants and strains using the Vienna {\it ab-initio} Simulation Package (VASP) version 4.6 \cite{VASP1}.  All calculations were performed with the Perdew-Burke-Ernzenhof generalized gradient corrections \cite{PBE}, and while we did investigate the effect of unpaired spins, Ca was found to be non-magnetic. As such, the results reported here are for spin-restricted PBE calculations. Forces for the calculation of phonons were found using 5x5x5 supercells in VASP.  Frozen phonon calculations were performed using fropho \cite{fropho}.

Calcium at zero pressure and temperatures below 721 K crystallizes into an fcc phase. At 300 K it remains fcc up to a pressure of 195 kbar, at which point it transforms to bcc \cite{Olijnyk}.   In this study, we focus attention on a region around the  low pressure fcc phase and find that it contains an isostructural transition.

\begin{figure*}
\includegraphics{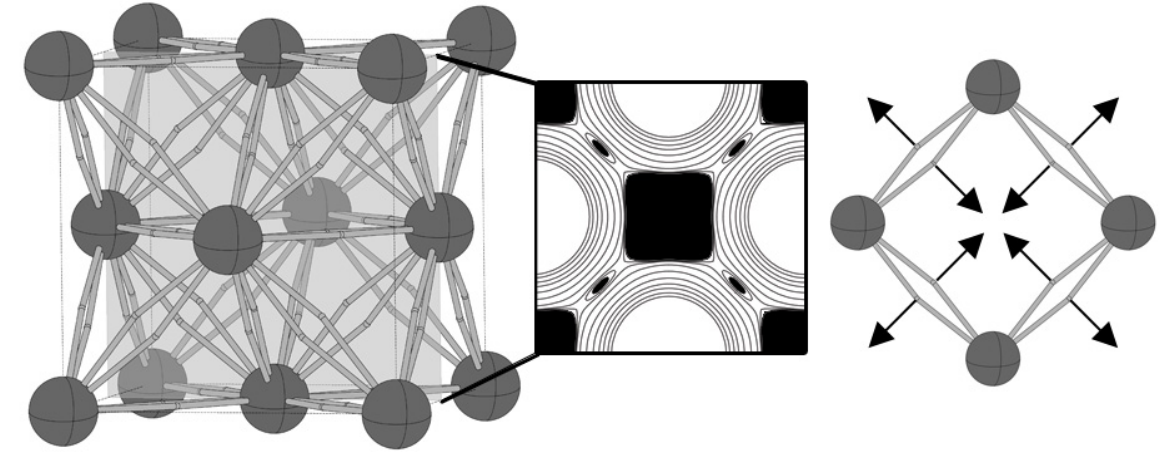} 
\caption{\label{fig:fig1} Atoms and bond paths in the low pressure topology of fcc Ca are shown left. Contours of the total charge density are shown for the (200) plane (center). The shaded regions represent minima in $\rho(\vec{r})$. The bifurcated bond paths and arrows indicating their motion under pressure are shown far right.}
\end{figure*}

Most nonmagnetic, monoatomic fcc  metals share the ground state charge density topology of the prototype, Cu \cite{JonesActa}, with each atom connected to its 12 nearest neighbors by single bond paths. The ground state charge density of Ca, however, is characterized by bifurcated bond paths, where bond point pairs are separated by a (3,+1) CP (Fig.~\ref{fig:fig1}). Like  the prototype, however, there is a (3,+3) CP at the center of the  octahedral hole. These features can be seen in the plot of $\rho(\vec{r})$ in the (200) plane shown in the center of Fig.~\ref{fig:fig1}. The position of the plane in the unit cell is shown in the left of Fig.~\ref{fig:fig1}. Minima in this plot are shaded black, while the maxima are unshaded. Ca atoms lie on the edges of the plot. 

As the pressure is increased towards the fcc-bcc transition pressure of 195 kbar, the bifurcated bond paths separate by moving towards the nearest octahedral cage CP. The four bond points nearest the octahedral hole in the center of the plot move towards that cage CP, while the other bond points move in the opposite directions, towards the octahedral holes at the corners of the plot (shown in the right of Fig.~\ref{fig:fig1}).  By symmetry, all of the CPs move at the same rate. Thus, the four bond points nearest the octahedral cage point in the center of Fig.~\ref{fig:fig1} will meet the cage point simultaneously, along with the four bond points on the two perpendicular planes, (020) and (002). All 12 CPs coalesce with the octahedral cage point when the external pressure reaches $\simeq$ 79 kbar \footnote{A separate topology change may occur at 24 kbar, transforming the bond points into nonnuclear maxima. The nature of this topological catastrophe appears to be different from the one being investigated and will be the subject of future research. However, for this work the type of critical point moving towards the octahedral hole is unimportant.}.  This topological catastrophe produces a non-nuclear maximum at the center of the fcc unit cell, the octahedral hole. 

\begin{figure}
\includegraphics{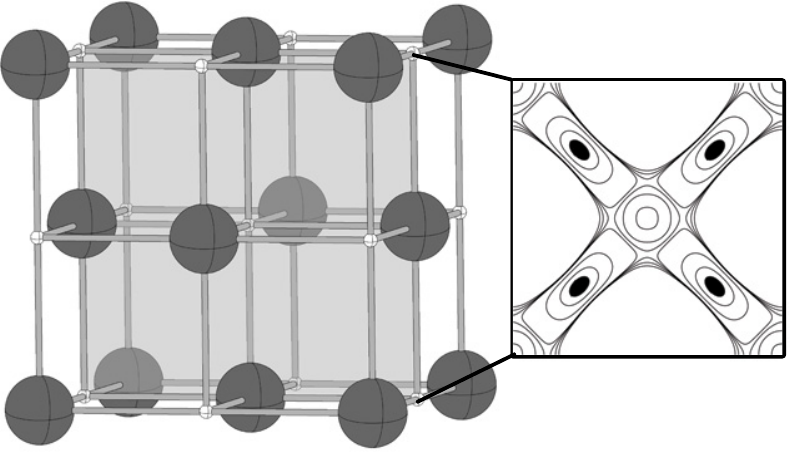} 
\caption{\label{fig:fig2} The atoms and bond paths in the high pressure topology of fcc Ca (left). Contours of the total charge density are shown for the (200) plane (right). The shaded regions represent local minima in $\rho(\vec{r})$.}
\end{figure}

The maxima and bond paths for this high pressure topology are shown in Fig.~\ref{fig:fig2}. Here, the larger gray spheres represent the Ca atoms, while the smaller white ones represent the non-nuclear maxima. Inspection of this figure reveals that all Ca atoms are now bound to the 12 nearest neighbor and 6 second neighbor atoms through the non-nuclear maxima located in the octahedral holes of the fcc lattice. The contour plot in the right of Fig.~\ref{fig:fig2} shows the contours of  $\rho(\vec{r})$ in the (200) plane. This is the same crystallographic plane shown in Fig.~\ref{fig:fig1}, with the same shading for minima and maxima. Comparing the two contour plots, it is clear that the CP in the octahedral hole has changed from a minimum to maximum while the CP on the atom-atom line remains a ring point. There are no further changes in the topology as the pressure is increased to 195 kbar.

A key result of this work is the discovery of a pressure-induced change to the topology of $\rho(\vec{r})$ that is not accompanied by a change in the crystallographic structure of fcc Ca.  This transition occurs at a pressure of $\simeq$ 79 kbar when the lattice constant is $\simeq$5.0 \AA.  The transition is the result of a topological catastrophe, through which the cage point at the center of an octahedral hole is transformed to a non-nuclear maximum.  

Because there is an abrupt change to calcium's charge density topology, we expect that properties measuring the response of $\rho(\vec{r})$ to an external perturbation will also change through this transition (e.g. the elastic properties of the material). To begin, we investigated the material's phonons, and while most stiffen with increasing pressure, the transverse shear wave propagating in the $[ 110 ]$ direction with polarization in the $[\bar{1} 1 0]$ direction softens after the topological catastrophe.  The stiffness of this mode is  given by the C$'$ single crystal elastic modulus.  

\begin{figure}
\includegraphics[scale=0.6]{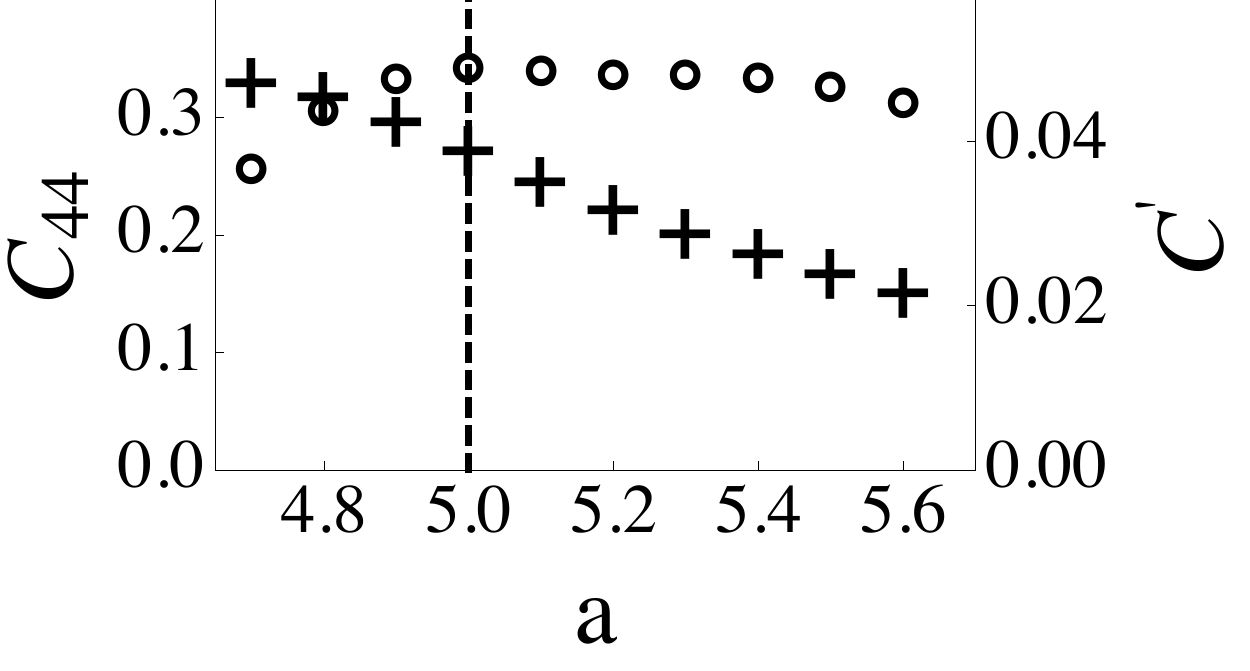} 
\caption{\label{fig:fig3} Plot of C$_{44}$ (crosses) and C$'$ (circles) in Mbar as a function of pressure in kbar. The C$'$ points are also connected by a dashed line to help guide the eye, and the lattice constant corresponding to the topological change is indicated with a vertical dashed line.}
\end{figure}

\begin{figure}
\includegraphics{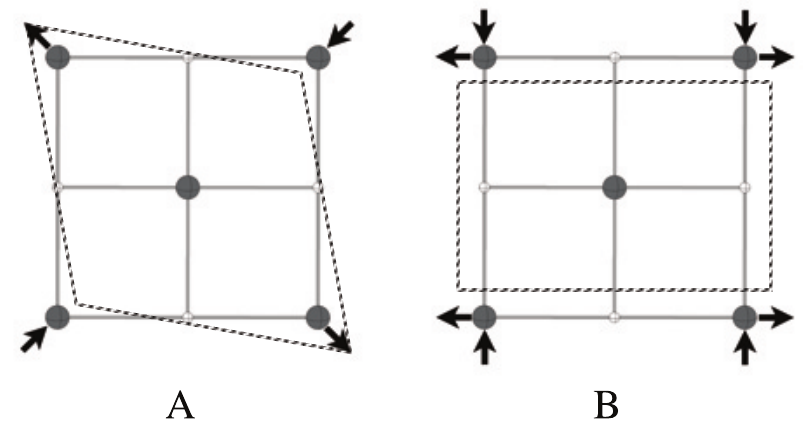} 
\caption{\label{fig:figstrain} Strains with only non-vanishing (A) $\epsilon_{xy}$  and  (B) $\epsilon_{xx}=-\epsilon_{yy}$. The dashed line indicates the strained square. Non-nuclear maxima in the high-pressure phase are denoted by open circles, while filled circles denote Ca atoms.}
\end{figure}

To understand the softening, consider both single crystal elastic moduli, C$_{44}$ and C$'$. These were calculated using VASP by employing a finite displacement approximation \cite{Cohen}. At the equilibrium lattice constant, these calculated single crystal elastic constants are in good agreement with experimental values. We found 0.046 Mbar and 0.16 Mbar for C$'$  and C$_{44}$ respectively, compared to the experimental values of 0.047 and 0.16 \cite{Stassis}. In the low pressure phase, both C$_{44}$ and C$'$ increase with applied pressure, while in the high pressure phase C$_{44}$ increases with applied pressure and C$'$ decreases (Fig.~\ref{fig:fig3}), with C$'$ achieving its  maximum value at the topological catastrophe.  This can be understood in terms of screening of the nuclear charges.  The strains resisted by C$_{44}$ and C$'$ are labeled  A and B respectively in Fig.~\ref{fig:figstrain}.  Strain A has non-vanishing $\epsilon_{xy}$, while strain B has $\epsilon_{xx}=-\epsilon_{yy}$.     Non-nuclear maxima in the high-pressure phase screen the restoring force for strain B and soften the elastic constant $C'$, while no such screening is present to soften C$_{44}$ .

We have found that the  topological catastrophe of $\rho(\vec{r})$ is the result of changes in the occupancy of calcium's one electron orbitals. Specifically,  calculations on  55 atom clusters of fcc Ca at various lattice constants using the Amsterdam Density Functional (ADF) package \cite{ADF} revealed that at the lattice constant corresponding to the catastrophe, filling an E$_g$ orbital comes at the expense of filling a T$_{2g}$ orbital. The latter is composed of s character on the second coordination sphere, while the former contains d$_{x^2-y^2}$ and d$_{z^2}$ character on the central atom, which contributes density to the octahedral hole of the fcc lattice. When the E$_g$ is filled, the extra density in the octahedral holes due to the anti-nodes on the d$_{x^2-y^2}$, d$_{z^2}$ orbitals  causes the topological change of $\rho(\vec{r})$.  

We propose using the distance {\it s} between the bond CPs and the cage CP as a convenient order parameter for the transition.  As the pressure is increased towards the critical pressure, P$_c$, the CPs around the octahedral hole move towards the center of the hole, as shown in Fig.~\ref{fig:fig1}, until they coalesce at the quantum critical point (QCP) with P=P$_c$ where {\it s} vanishes.  On the basis of numerical studies near the QCP, we show that {\it s} follows the mean-field scaling law with reduced pressure $p\equiv|P-P_c|/P_c$.  

\begin{figure}
\includegraphics[scale=0.5]{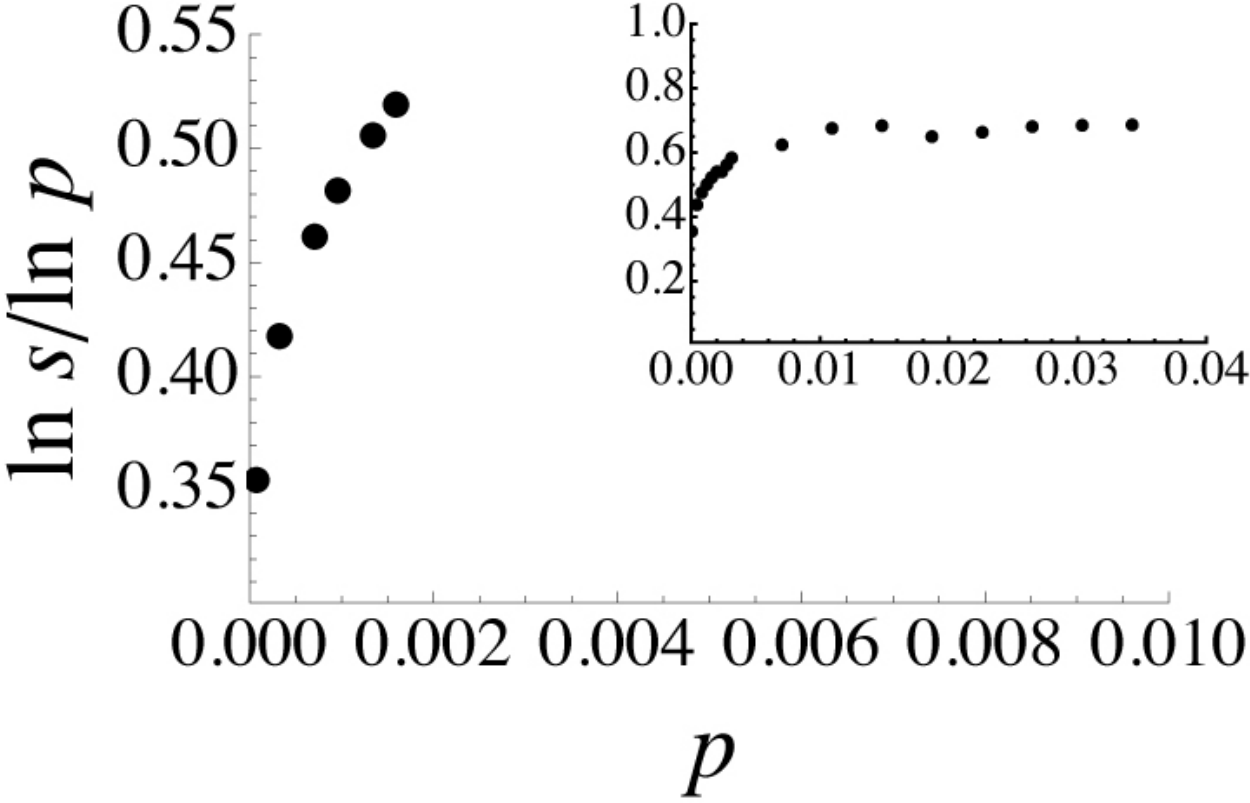} 
\caption{\label{fig:figorder} Plot of $\ln{\it s}/\ln p$ vs. ${p}$.  Extrapolation to $p=0$ gives a limiting value of $1/\delta=0.33$, consistent with the mean-field scaling law. }
\end{figure}

Using the well-known mapping between quantum systems at zero temperature in $d$-dimensions and classical systems in $d+1$-dimensions, we anticipate that in 3d a scalar order parameter near the QCP should scale with reduced pressure as ${\it s}\sim p^{1\over\delta}$ with the mean-field critical exponent $\delta=3$ taken.  Fig.~\ref{fig:figorder} shows DFT results of $\ln p/\ln {\it s}$ near the QCP.  

A very fine charge density mesh was required to achieve the desired accuracy.  However, our  results remained unchanged at grid densities beyond  200 points per Angstrom. As such, the results reported here were calculated using 200 points per Angstrom mesh. 
We find that $1/\delta$ reaches a value of 0.35 at {\it p} $\simeq$ 8e-5, and  converges to 0.33 upon linear extrapolation to {\it p} = 0. (Quadratic and cubic extrapolations were also performed using the 6 points nearest {\it p} = 0, leading to 0.34 and 0.33, respectively.)

In summary, we have used the topology of $\rho(\vec{r})$ to predict an isostructural phase transition in fcc calcium. We found that fcc calcium adopts different topologies of $\rho(\vec{r})$ when the applied hydrostatic pressure is above and below 79 kbar. We predict that this topological change in the charge density softens the C$'$ elastic modulus of fcc Ca, while C$_{44}$ remains unchanged.  We propose an order parameter based on applying Morse theory to the charge density, and we show that near the critical point it follows the appropriate mean-field scaling law with reduced pressure.  We gratefully acknowledge support of this work by ONR under grant N00014-10-1-0838 and NSF under  DMR-0814377.

\bibliography{Refs1}

\end{document}